\documentclass[aps,preprint,amsmath,amssymb]{revtex4}
\usepackage{graphicx}

\begin{document}

\title
{\Large \bf Neutrino Masses, Leptogenesis and Decaying Dark Matter}

\author{ \bf  Chuan-Hung Chen$^{1,2}$\footnote{Email:
physchen@mail.ncku.edu.tw},  Chao-Qiang Geng$^{3}$\footnote{Email:
geng@phys.nthu.edu.tw} and
Dmitry~V.~Zhuridov$^{3}$\footnote{Email:
zhuridov@phys.nthu.edu.tw}
 }

\affiliation{ $^{1}$Department of Physics, National Cheng-Kung
University, Tainan 701, Taiwan \\
$^{2}$National Center for Theoretical Sciences, Hsinchu 300, Taiwan
\\
$^{3}$Department of Physics, National Tsing-Hua University,
Hsinchu 300, Taiwan
 }

\date{\today}

\begin{abstract}
We study a simple extension of the standard model to simultaneously
explain neutrino masses, dark matter and the matter-antimatter
asymmetry of the Universe. In our model, the baryon asymmetry is
achieved by the leptogenesis mechanism, while the decaying dark
matter with the lifetime of $\mathcal{O}(10^{26}\,s)$  provides a
natural solution to the  electron and positron excesses in  Fermi
and PAMELA satellite experiments. In particular, we emphasize that
our model is sensitive to the structure at the endpoint around 1~TeV
of the Fermi data. In addition, some of  new  particles proposed  in
the model are within the reach at the near future colliders, such as
the Large Hadron Collider.
\end{abstract}

\maketitle

The observed neutrino oscillations and  matter-antimatter asymmetry
as well as the evidence for dark matter (DM)~\cite{PDG} clearly
imply physics beyond the standard model (SM). Recently,
PAMELA~\cite{PAMELA} and ATIC~\cite{ATIC} cosmic-ray measurements
show the positron/electron excesses above the calculated backgrounds
for the energy of  $\mathcal{O}(100)$ GeV. These data are consistent
with the measurements of the high energy electrons and positrons
fluxes in the cosmic ray spectra by PPB-BETS~\cite{PPB-BETS},
HEAT~\cite{HEAT}, AMS~\cite{AMSCollab} and HESS~\cite{HESS,HESSnew}.
Very recently, a more precise data by the Fermi LAT
collaboration~\cite{FermiLAT} also indicates some enhancements in
the electrons + positrons flux in the $100-1000$ GeV energy range.
However, the Fermi's result is in conflict with the large excess in
flux around 500 GeV range  by ATIC. Similar conclusion has also been
given by HESS based on the low energy data~\cite{HESSnew}. In this
study, we will concentrate on the combined data of PAMELA and Fermi
(PF) without fitting that of ATIC.

To explain the PAMELA/ATIC data, there have been many possible
mechanisms, such as DM decays~\cite{DMdecay,0901.2681}, DM
annihilations~\cite{DMannih} and astrophysical
sources~\cite{pulsars}, while recent studies related to the PF data
without the ATIC one can be found in
Refs.~\cite{splitSUSY,Fermi-studies}.
In this paper, we would like to explore the possibility of
connecting the neutrino masses to the dark matter problem as well as
the baryon asymmetry of the Universe (BAU). In particular, we would like to pay attentions
to models which can be tested directly by the future high energy
colliders, such as the Large Hadron Collider~\cite{LHC} and Linear
Collider (LC)~\cite{LC}.

We introduce three new neutral leptons $N_i$ ($i=1,2$) and $N$ with
the masses of $M_i$ and $M$, and two new doublet scalars $\zeta$ and
$\eta$ with zero VEVs and the masses of $M_\zeta$
and $M_{\eta}$, respectively, in the SM. These new particles have
non-trivial
transformation properties under the
two discrete symmetries $Z_2$ and $Z_2^\prime$ as listed in
Table~\ref{T1}, whereas the corresponding SM particles are trivial.
\begin{table}[htb]
\begin{center}
\caption{ Transformations of the new
particles
under  the discrete symmetries of $Z_2$  and $Z_2^\prime$.}\label{T1}
\begin{tabular}{|c|c|c|c|c|}
  \hline
\ Particle & $\ \zeta$ & \ $\eta$ & \ $N_i$ & \ $N$ \\
\hline
\ $Z_2$ & \ $-$ & \ $+$ & \ $-$ & \ $+$ \\
\hline
\ $Z_2^\prime$ & \ $+$ & \ $-$ & \ $+$ & \ $-$ \\
  \hline
\end{tabular}
\end{center}
\end{table}

The relevant Majorana mass  terms and Yukawa couplings as well as
the soft breaking term involving the new particles can be written as
\begin{eqnarray}
    \frac{M_{ij}}{2}N_i^TCN_j + \frac{M}{2}N^TCN + y_{ij}\bar L_i\zeta N_{j} + y_{i}^\prime\bar L_i\eta N + \mu^2\eta^\dag\zeta + {\rm
    H.c.},
\end{eqnarray}
where $i$ and $j$ are the flavor indexes and $L_i$ are the lepton
doublets in the SM.
We note that the soft breaking term in Eq. (1) breaks the two
discrete symmetries to a diagonal one.
% $Z_2^{\prime\prime}$.
 In our
study, we will assume the mass hierarchies  of
$M_{\zeta^0}<M_{\zeta}< M_1<M_2$ and $M_{\zeta}< M<M_\eta$.
%%%%%%%%%%%
We will demonstrate that the leptogenesis is achieved by $N_1$ decays, while $N$ is
the decaying dark matter.
%%%%%%%%%%%%%

\begin{figure}[ht]
\centering
\includegraphics*[width=2.8in]{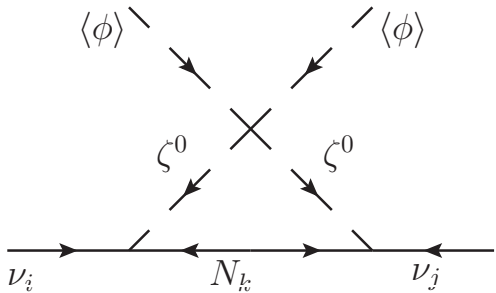}
\caption{Neutrino masses generated at one-loop level.}
 \label{Fig:nu masses}
\end{figure}

The neutrino masses are generated by the one-loop
diagram in Fig.~\ref{Fig:nu masses} as proposed in Ref.~\cite{Ma} due to the
quartic scalar interaction of
\begin{eqnarray}
    \frac{\lambda}{2}(\phi^\dag\zeta)^2 + {\rm
    H.c.}\,,
\end{eqnarray}
where $\phi$ is the SM Higgs boson.
A simple formula for the neutrino masses is found to be~\cite{Ma,Gu_Sarkar}
\begin{eqnarray}
    (m_\nu)_{ij} = \frac{\mathcal{O}(\lambda)}{16\pi^2} \sum_{k=1}^2
    \frac{y_{ik}y_{jk}}{M_k}v^2
\end{eqnarray}
with the SM Higgs VEV of $v\simeq 174$~GeV.
For the parameter set of $\lambda =
\mathcal{O}(10^{-4})$, $y_{ij} = \mathcal{O}(10^{-3})$ and $M_i =
\mathcal{O}(100~{\rm GeV} - 10~{\rm TeV})$,
we obtain $m_\nu = \mathcal{O}(0.01-0.1~{\rm eV})$,  consistent with the current neutrino data.
Similarly to the minimal seesaw model with two right-handed
neutrinos~\cite{twoNR}, our model contains
one
%one of the observed neutrinos is
massless neutrino with
%which allows
only normal or inverted hierarchy of the neutrino
masses. However, the extended model with three $N_i$ could still has  the
possibility of the quasi-degenerate neutrino masses.

\begin{figure}[ht]
\centering
\includegraphics*[width=6.2in]{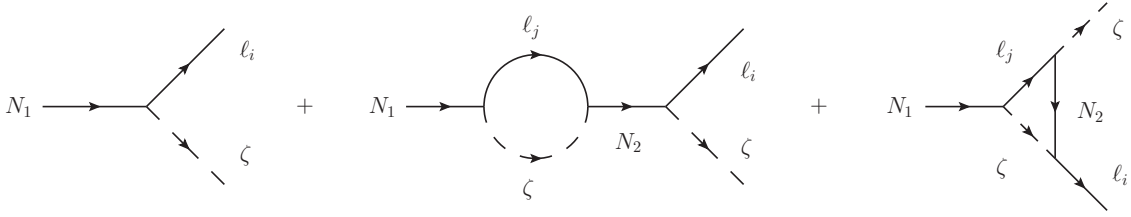}
\caption{Tree level
and one-loop diagrams for $N_1\to\ell_i\zeta$.
}
 \label{Fig:LG}
\end{figure}

The leptogenesis mechanism in our paper is the same as that in Ref.~\cite{Gu_Sarkar},
provided by $N_i$ and $\zeta$.
The relevant diagrams for the leptogenesis are shown in Fig.~\ref{Fig:LG}.
The decay width of $N_1$ is given by
\begin{eqnarray}
    \Gamma(N_1\to e^\mp\zeta^\pm) &=&
    \frac{(y^\dag y)_{11}}{16\pi}M_1r^2
\end{eqnarray}
with $r=1-M_\zeta^2/M_1^2$.
The out-of-equilibrium condition requires $r \sim 10^{-4}$~\cite{Gu_Sarkar}.
%%%%%%%%%%%%%%%%%%%
We remark that this small value of $r$ implies some degeneracy
between $M_1$ and $M_\zeta$. However, it may be avoided by including
another new doublet with a soft breaking term similar to the
discussion on the decaying dark matter discussed later.
%%%%%%%%%%%%%%%%%%%

In the case of $3M_1<M_2$, the $CP$ violating parameter in the
leptogenesis is given by
\begin{eqnarray}
    \varepsilon \simeq -\frac{3}{16\pi} \frac{1}{(y^\dag y)_{11}}
    {\rm Im} \left[ (y^\dag y)_{12}^2 \right]
    \frac{M_1}{M_2}.
\end{eqnarray}
By using $y_{ij} = \mathcal{O}(10^{-3})$ and $M_i =
\mathcal{O}(100~{\rm GeV} - 1~{\rm TeV})$ and assuming a maximal CP
phase, the net BAU can be obtained as
\begin{eqnarray}
    \frac{n_B}{s} \simeq -\frac{1}{15}\frac{\varepsilon}{g_*} \simeq
    10^{-10},
\end{eqnarray}
where $g_* \simeq 100$ is the relativistic degrees of freedom.
The decays of $\zeta^+ \to \zeta^0\ell^+\nu$ and $\zeta^- \to
\zeta^0\ell^-\bar\nu$  help to avoid the dangerous relics from
the singly-charged component of $\zeta$, while $\zeta^0$ may provide
only a sub-dominant component of the DM due to the annihilation into
gauge bosons~\cite{inertHiggs}.

\begin{figure}[ht]
\centering
\includegraphics*[width=2.8in]{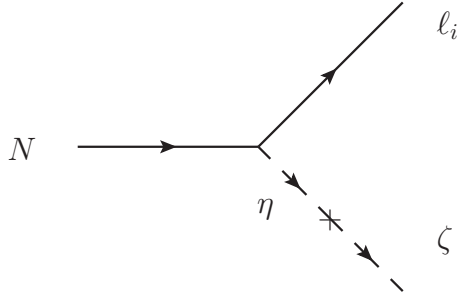}
\caption{Diagram for the DM decay.}
 \label{Fig:PAMELA-ATIC}
\end{figure}

The diagram for the DM decay is shown in Fig.~\ref{Fig:PAMELA-ATIC}.
The decay width of $N$
%\to \ell_i^\pm\zeta^\mp$
is given by
\begin{eqnarray}
    \Gamma_i = \frac{|y_i^\prime|^2}{4\pi}\left(\frac{|\mu|}{M_\eta}\right)^4
    \frac{M_{-}^2}{M}, %\simeq \frac{|y_i^\prime|^2}{32\pi}
    %\left(\frac{|\mu|}{M_\eta}\right)^4M,
\end{eqnarray}
where we have used the definition of
\begin{eqnarray}
    M_\pm=\frac{M^2\pm M_\zeta^2}{2M}.
\end{eqnarray}
By neglecting the effects of the off-shell $\zeta^\pm$, the lifetime
of $N$ is
%can be derived to be
\begin{eqnarray}
    \tau_{N} = \frac{1}{4\sum_{i}\Gamma_{i}} %= \frac{32\pi A^4}{3M}
    %\frac{128(2\pi)^3}{3}\frac{M_0^2M^5}{(\Delta
    %M_{1}^2)^4},
    = \frac{\pi A^4M}{M_-^2}
\end{eqnarray}
with
\begin{eqnarray}
    A = \frac{M_\eta}{|\mu|(\sum_{i}|y_{i}^\prime|^2)^{1/4}}\,,
\end{eqnarray}
where we have included both charged and neutral modes. By taking
$y_{i}^\prime=\mathcal{O}(10^{-4})$, $\mu=\mathcal{O}(1\,{\rm keV})$
and $M_\eta=\mathcal{O}(100$~TeV), one gets $A\sim 10^{13}$, leading
to $\tau_N\sim 10^{26}$~s with $M_\zeta\sim 0.5$~TeV and $M\sim
2$~TeV.

The normalized energy spectrum of the electron/positron for
$N\to e^\mp\zeta^\pm$  can be written as
\begin{eqnarray}
    \frac{dN_e}{dE} = 2\delta(M_{-}-E),
\end{eqnarray}
while that for the  decaying
chain of $N\to \zeta^\pm\mu^\mp(\to e^\mp2\nu)$ is~\cite{splitSUSY}
\begin{eqnarray}
    \frac{dN_{\mu e}}{dE} =
    \frac{4}{3M_{-}}\left[\left(x^3-1\right) -
    \frac{9}{4}\left(x^2-1\right)\right]
\end{eqnarray}
with $x=E/M_-$ and $0<E<M_-$.
For the electron/positron produced in the $\zeta^\mp\to
e^\mp\nu\zeta^0$ subprocesses via exchanges of $W^\mp$ bosons, we
have
\begin{eqnarray}
    \frac{dN_{\zeta e}}{dE} &=&
    \left[ \int_0^{E_{max}} dE \frac{d\tilde N_{\zeta e}}{dE}
    \right]^{-1} \frac{d\tilde N_{\zeta e}}{dE},
%\begin{eqnarray}
%    \frac{dN_{\zeta e}}{dE} = C \int_{M_\zeta}^{M} dE_\zeta \frac{d\tilde N_\zeta}{dE_\zeta}
%    \frac{d\tilde N_{\zeta e}}{dE}, \label{Eq:zeta distribution}
\end{eqnarray}
where
\begin{eqnarray}
    E_{max}&=&\frac{M\left(M_{\zeta}^2-M_{\zeta^0}^2\right)}{2M_\zeta^2},\\
    \frac{d\tilde N_{\zeta e}}{dE} &=& %\left[ (m_{\nu e}^2)_+^3 - (m_{\nu
    %e}^2)_-^3 \right]  -  \frac{3M_W^2}{2} \left[
    (m_{\nu e}^2)_+^2
    - (m_{\nu e}^2)_-^2 %\right]
\end{eqnarray}
with
\begin{eqnarray}
    &&(m_{\nu e}^2)_\pm = (E_1^*+E_3^*)^2 - (|E_1^*|\mp
    |E_3^*|)^2,\\
    &&E_1^* = \frac{m_{\nu\zeta^0}^2-M_{\zeta^0}^2}{2m_{\nu\zeta^0}}, \quad  E_3^* =
    \frac{M_\zeta^2-m_{\nu\zeta^0}^2}{2m_{\nu\zeta^0}}, \\
    &&m_{\nu\zeta^0}^2 = M_\zeta^2\left(1-2E/M\right).
\end{eqnarray}
Note that we will concentrate on the case in which $\Delta
M_\zeta\equiv M_\zeta-M_{\zeta^0}$ is small to forbid the hadronic
decay modes of $W$ bosons. The normalized resultant energy spectrum
of the electron/positron from the DM decays can be written as
\begin{eqnarray}
    \frac{dN}{dE} = \frac{1}{2+\varepsilon} \left[ \frac{dN_e}{dE}
    + \varepsilon\frac{dN_{\mu e}}{dE}%\theta(M_--E)
    + \frac{dN_{\zeta e}}{dE}\theta\left(E_{max}-E\right)
    %+ \varepsilon\frac{dN_{\zeta\mu e}}{dE} \theta\left(\frac{M_{eff}}{4}-E\right)
    \right],
\end{eqnarray}
where $\varepsilon=|y_\mu^\prime|^2/|y_e^\prime|^2$ and we have assumed
100\% rate for the electron channel of the $\zeta$ decay.
We remark that we have neglected the tau-lepton effect in this study, but it can
be included straightforwardly.

%%% Results %%%%%%%%%%%%%%%%%%%%%%%%%%%%%%%%%%%%%%%%%%%%%%%%%%%%%%%%%%%%%%%%%%%%%%%%%%%%%%%%%%%%%%%%%%%%%%%%%%%%%%%%%%%%%%%%%%%%%%%%%%%%%

The DM component of the primary electron/positron flux is given
by~\cite{Ibarra_Tran,DMdecay}
\begin{eqnarray}
    \Phi_{e}^{DM}(E)=\frac{c}{4\pi M\tau_{N}}\int\limits_0^{M_-}dE^\prime G(E,E^\prime)\frac{dN}{dE^\prime},
\end{eqnarray}
where $E$ is in units of GeV and $c$ is the speed of light. All
the information about astrophysics is encoded in the Green
function of $G(E,E^\prime)$,
%approximately
given by
\begin{eqnarray}
    G(E,E^\prime) \simeq \frac{10^{16}}{E^2}\exp[a+b(E^{\delta-1}-E^{\prime\delta-1})]\theta(E^\prime-E) \quad [{\rm cm}^{-3}{\rm
    s}],
\end{eqnarray}
where the normalization is adjusted to yield a local halo density
$\rho_\odot\sim 1$~GeV\,cm$^{-3}$~\cite{Cirelli}. We use the
coefficients of $a=-1.0203$ and $b=-1.4493$~\cite{Ibarra_Tran} for
the spherically symmetric Navarro, Frenk and White density profile
of the DM in our Galaxy~\cite{NFW} and the diffusion parameter
$\delta=0.70$ for the MED propagation model~\cite{MED}, which is
consistent with the observed Boron-to-Carbon ratio~\cite{BtoC}.
Here, we have not taken account of the charge-sign dependent solar
modulation~\cite{Blatz} as well as  other astrophysical
uncertainties~\cite{MED}, which could be significant in the energies
below 10~GeV.
%.
The total electron and positron fluxes are
\begin{eqnarray}
    \Phi_{e^-}&=&\kappa\Phi^{prim}_{e^-}+\Phi^{DM}_{e^-}+\Phi^{sec}_{e^-},
    \nonumber\\
    \Phi_{e^+}&=&\Phi^{DM}_{e^+}+\Phi^{sec}_{e^+},
\end{eqnarray}
respectively, where $\Phi^{prim}_{e^-}$ is a primary astrophysical
component, presumably originated  from supernova remnants,
$\Phi_{e^{-(+)}}^{DM}$ is an exotic primary component from the DM
decays, $\Phi_{e^{-(+)}}^{sec}$ is a secondary component from the
spallation of cosmic rays on the interstellar medium, and $\kappa$
is a free parameter about 1 to fit the data when there is no DM
primary source. We choose $\kappa=0.7$ to insure the flux calculation
to be consistent with the  data. For the background fluxes, we will
use the parameterizations obtained in Refs.~\cite{Blatz,Moskalenko},
given by
\begin{eqnarray}\label{Eq_e-}
    \Phi_{e^-}^{prim}(E)&=&\frac{0.16E^{-1.1}}{1+11E^{0.9}+3.2E^{2.15}} \quad
    [{\rm GeV}^{-1}{\rm cm}^{-2}{\rm s}^{-1}{\rm sr}^{-1}],\\
    \Phi_{e^-}^{sec}(E)&=&\frac{0.7E^{0.7}}{1+110E^{1.5}+600E^{2.9}+580E^{4.2}} \quad
    [{\rm GeV}^{-1}{\rm cm}^{-2}{\rm s}^{-1}{\rm sr}^{-1}],\\
    \Phi_{e^+}^{sec}(E)&=&\frac{4.5E^{0.7}}{1+650E^{2.3}+1500E^{4.2}} \quad
    [{\rm GeV}^{-1}{\rm cm}^{-2}{\rm s}^{-1}{\rm sr}^{-1}],
    \label{Eq_e+}
\end{eqnarray}
where $E$ is in units of GeV.

\begin{figure}[ht]
\centering
\includegraphics*[width=3in]{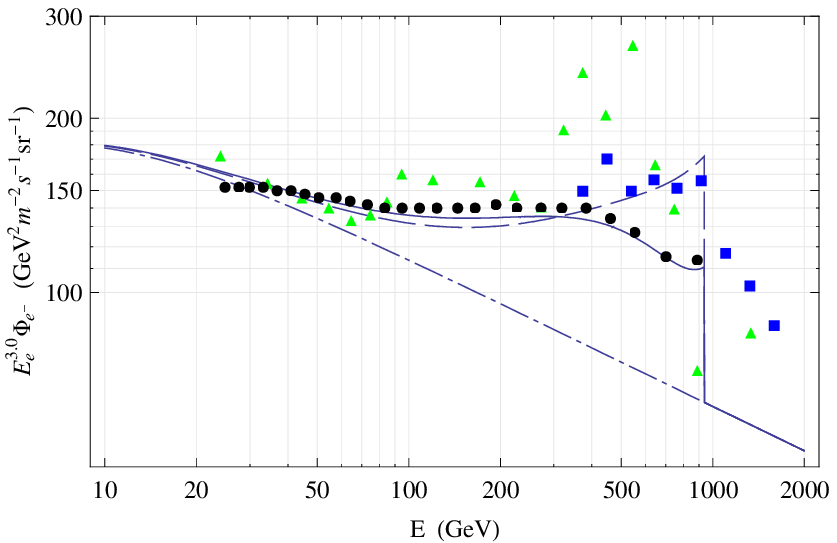}
\includegraphics*[width=3in]{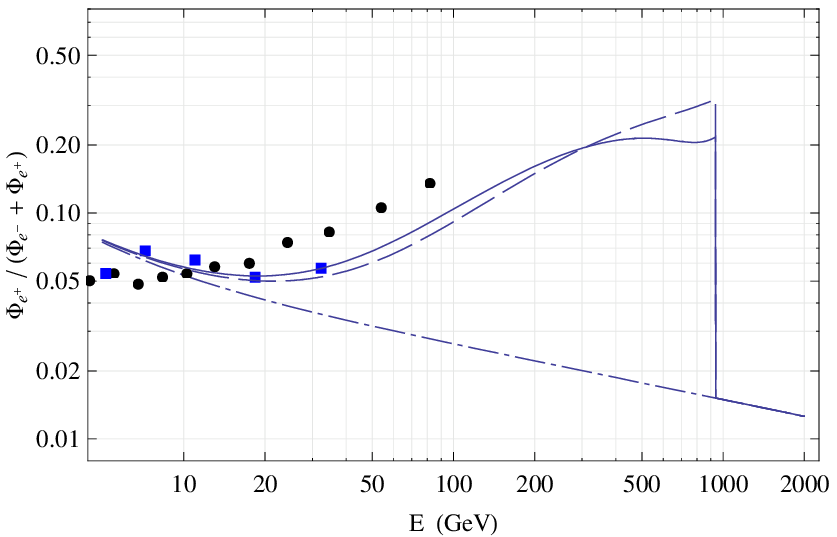}
\caption{ Electron + positron energy spectrum (left) and positron
fraction (right) of the DM decays with $\tau_N=2.5\times10^{26}$~s,
$M=2$~TeV, $M_\zeta=500$~GeV and $\Delta M_\zeta=1$~GeV, where
$\varepsilon=1$ (5) is represented by dashed (solid) lines, black
points and blue rectangles stand for the observations of Fermi and
HESS (left) and PAMELA and HEAT (right), and green triangles and
dot-dashed lines correspond to the ATIC and backgrounds,
respectively.}
 \label{Fig:MED}
\end{figure}
\begin{figure}[ht]
\centering
\includegraphics*[width=3in]{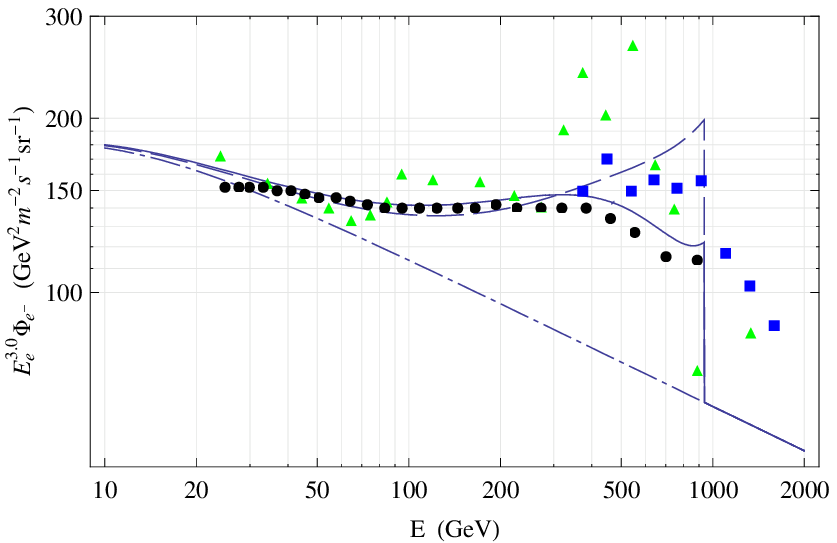}
\includegraphics*[width=3in]{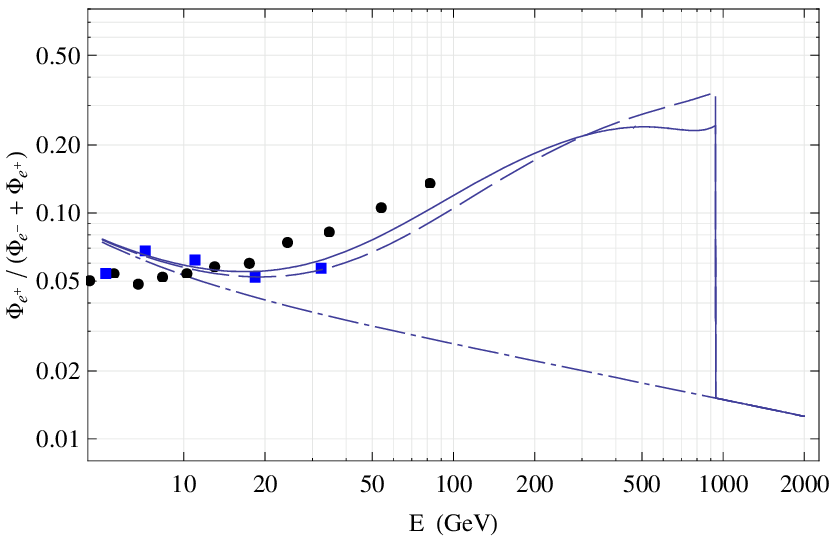}
\caption{ Legend is the same as Fig.~4 but with
$\tau_N=2\times10^{26}$~s.}
 \label{Fig:MEDtau2}
\end{figure}

In Fig.~\ref{Fig:MED} [\ref{Fig:MEDtau2}], we show the electron plus
positron energy spectrum (left) and the positron fraction (right) of
the DM decays for $\tau_N=2.5\times10^{26}\,[2\times10^{26}]$~s,
$M=2$~TeV, $M_\zeta=500$~GeV, $\Delta M_\zeta=1$~GeV and
$\varepsilon=1$ and 5, respectively, where the backgrounds are
represented by dot-dashed lines. The electron + positron flux is
multiplied by $E^3$ to compensate the roughly $E^{-3}$ falling of
the flux. From the figures, we see
that the model with the enhanced muon effects is in good agreement
with the Fermi, HESS, PAMELA and HEAT data. In particular, the
energy spectrums in Figs.~\ref{Fig:MED} (left) and \ref{Fig:MEDtau2}
(left) with $\varepsilon=5$ perfectly matches the Fermi's  result.
It is worth to mention that the results for the energy spectrum and
positron fraction are not significantly dependent on the $\zeta$
mass in the wide range of $100~{\rm GeV}-1~{\rm TeV}$ besides the
end point moving to a higher energy for a smaller $M_\zeta$.
%for a smaller lifetime, e.g.
%$\tau_N=2\times10^{26}$~s, one could fit the PAMELA and ATIC data
%very well  as those in Refs.~\cite{DMdecay,0901.2681} by ignoring
%the data of Fermi and HESS.
%In Figs.~\ref{Fig:MED} and \ref{Fig:MED150GeV}, the electron +
%positron energy spectrum and positron fraction have two peaks due to
%the two independent mass splittings among  $M$, $M_\zeta$ and
%$M_{\zeta^0}$. The lower energy peak is mild due to the $\zeta^\pm$
%three body decays, while the higher one around 1~TeV is softened by
%the muon effect. It is interesting to note that the structures of
%the energy spectrum and positron fraction are more significant for a
%smaller muon contribution. In particular, we emphasize that future
%precision  cosmic rays data
%from Fermi or other experiments
%could be important to prove or disprove the interesting structure
%around 1000~GeV.
We remark that for a
lighter DM particle, the drop in the electron flux occurs at a lower
energy compared to the Fermi  data.
%

%%%%%%%%%%%%%%%%%%%%%%%%%%%%%%%%%%%%%%%%%%%%%%%%%%%%

As for the collider signatures from the new particles in the
 model, there are possible pair productions of $\zeta$
directly by the SM gauge bosons~\cite{Ma}. However, the pair productions of
$N_i$ by the $e^+e^-$ annihilations through the $\zeta$
exchanges~\cite{Gu_Sarkar} are hard to be observed
at the near future LC due
to the small values of $y_{ij}=\mathcal{O}(10^{-3})$.
In addition,  although $N$ and
$\eta$  will escape the detection in the next generation of
colliders due to their weak couplings and heavy masses,
their properties can be tested by precise measurements of the
electron spectrum and positron fraction since
%t
 the signals are not sensitive to the propagation
models at the energies
higher than 400~GeV. In particular, future measurements of the positron fraction
at energies higher than 100~GeV can be crucial in testing the same
origin of the
Fermi and PAMELA electron and positron
excesses.
%

%%%%%%%%%%%%%%%%%%%%%%%%%%%%%%%%%%%%%%%%%%%%%%%%%%%%%

Finally, we remark that the neutral lepton $N$ could be produced copiously after the big bang
to become a typical unwanted relic. However, the inflation
dilutes $N$ away since its number density reduces exponentially.
The present abundance of $N$ may be generated by the $e^+e^-$
annihilation.

In conclusion, we have investigated a  relatively simple
 extension of the SM to generate the small neutrino masses at one-loop level
 and the observed BAU by
the leptogenesis mechanism.
Our model also contains
 the decaying dark matter
with the lifetime of $\mathcal{O}(10^{26}\,s)$,  which provides
a natural solution to the  electron and positron excesses  in
cosmic rays in  the energy range of  $100-1000$~GeV by Fermi and PAMELA.
It should be emphasized that the structure  at the endpoint around 1000~GeV of the Fermi data
is crucial to determine the muon effects in the dark matter decays.
More precise cosmic ray measurements around this energy range are clearly needed.
%In addition, some of  new  particles
%proposed  in the model are within the reach at the near future
%colliders, such as the Large Hadron Collider.

\section*{Acknowledgements}
We would like to thank Dr. Takeshi Araki for useful discussions. This
work is supported in part by the Boost Program of NTHU and the National Science Council of R.O.C.
under Grant Nos: NSC- 97-2112-M-006-001-MY3 and
NSC-95-2112-M-007-059-MY3.

%%%%%%%%%%%%%%%%%%%%%%%%%%%%%%%%%%%%%%%%%%%%%%%%%%%%%%%%%%%%%%%%%%%%%%%%%%%%%%%%%%%%%%%%%%%%%%%%%%%%%%%%%%%%%%%%%%%%%%%%%%%
%%%%%%%%%%%%%%%%%%%%%%%%%%%%%%%%%%%%%%%%%%%%%%%%%%%%%%%%%%%%%%%%%%%%%%%%%%%%%%%%%%%%%%%%%%%%%%%%%%%%%%%%%%%%%%%%%%%%%%%%%%%

\end{document}